\documentclass[superscriptaddress,prl,amsmath,amssymb,eufrak,aps,floatfix,twocolumn,notitlepage,longbibliography]{revtex4-1}

\newcommand{\rpd}[1]{\partial_t #1}

\usepackage{lipsum}
\usepackage{braket}
\usepackage{graphicx}
\usepackage{dcolumn}
\usepackage{enumitem}
\usepackage{bm}
\usepackage{amsmath}
\usepackage{amssymb}
\usepackage{mathtools}
\usepackage{subfigure}
\usepackage{color}
\usepackage{xcolor}
\usepackage{verbatim}
\usepackage{float}
\usepackage{appendix}
\usepackage{bbold}

\newcommand{\ev}[1]{\langle #1 \rangle} 
\newcommand{\hSig}[0]{\hat{\sigma}} 

\begin{document}

\preprint{APS/123-QED}

\title{Subradiant-to-Subradiant Phase Transition in the Bad Cavity Laser}

\author{Athreya Shankar}
\thanks{These authors contributed equally to this work. Corresponding author: A.S.; athreya.shankar@uibk.ac.at}
\affiliation{Center for Quantum Physics, Faculty of Mathematics, Computer Science and Physics, University of Innsbruck, Innsbruck A-6020, Austria}
\affiliation{Institute for Quantum Optics and Quantum Information, Austrian Academy of Sciences, Innsbruck A-6020, Austria}
\author{Jarrod T. Reilly}
\thanks{These authors contributed equally to this work. Corresponding author: A.S.; athreya.shankar@uibk.ac.at}
\affiliation{JILA, NIST, and Department of Physics, University of Colorado Boulder, Boulder, Colorado 80309, USA}
\author{Simon B. J\"ager}
\affiliation{JILA, NIST, and Department of Physics, University of Colorado Boulder, Boulder, Colorado 80309, USA}
\author{Murray J. Holland}
\affiliation{JILA, NIST, and Department of Physics, University of Colorado Boulder, Boulder, Colorado 80309, USA}

\date{\today}

\begin{abstract}
We show that the onset of steady-state superradiance in a bad cavity laser is preceded by a dissipative phase transition between two distinct phases of steady-state subradiance. The transition is marked by a non-analytic behavior of the cavity output power and the mean atomic inversion, as well as a discontinuity in the variance of the collective atomic inversion. In particular, for repump rates below a critical value, the cavity output power is strongly suppressed and does not increase with the atom number, while it scales linearly with atom number above this value. Remarkably, we find that the atoms are in a macroscopic entangled steady state near the critical region with a vanishing fraction of unentangled atoms in the large atom number limit.
\end{abstract}

\maketitle
\emph{Introduction.---} Progress in laser physics has revolutionized our day-to-day lives and the scope of experiments across the entire spectrum of scientific disciplines. At its core, the laser is a highly out-of-equilibrium system whose steady state is maintained via a balance of driving and dissipation. A typical laser model involves a collection of continuously pumped two-level atoms interacting with an electromagnetic field confined in a cavity with lossy mirrors. In particular, bad cavity lasers operate in a regime where the lifetime of the photon is short compared to the effective lifetime of the upper atomic level \cite{meiser2009PRL,meiser2010PRASteadyState,bohnet2012Nature}. Over the past decade, they have garnered significant attention because in these systems the sensitivity of the laser linewidth to cavity frequency fluctuations is strongly suppressed \cite{meiser2009PRL,liuPRL2020}. Furthermore, this narrow linewidth coexists in a regime where the emission amplitudes of the atoms can constructively interfere and give rise to superradiant emission. Apart from its promising technological potential, the superradiant regime has also been shown to host a variety of many-body phenomena such as synchronization \cite{Acebron:2005,Cube:2004,Zhu:2015,Heinrich:2011,xu2014PRL,xu2015PRL,weiner2017PRA,Bellomo:2017,patra2019PRA_1,patra2019PRA_2,tucker2018NJP}, collective cooling \cite{Domokos:2002,Black:2003,Chan:2003,xu2016PRL,jaeger2017PRA,Hotter:2019} and self-organization \cite{Domokos:2002,Ritsch:2013,Baumann2010,Arnold:2012,jaeger2019PRL,Jager:2020}. 

In contrast, the regime preceding the onset of superradiance has received far less attention, partly because within the framework of mean-field theory the atoms appear to be in a trivial unpolarized product state. Prior beyond-mean-field studies have only considered this regime in passing \cite{meiser2010PRASteadyState,meiser2010PRAIntensity} or for a small number of emitters \cite{temnov2009OptExp,auffeves2011NJP}, but have nevertheless demonstrated that the atoms populate collective dark states giving rise to steady-state subradiance. However, the physics in this regime and the stability of the highly-correlated quantum states remains poorly understood especially given the fact that this regime is complementary to the well studied and much anticipated steady-state superradiant regime.

\begin{figure}[!tb]
    \centering
    \includegraphics[width=\columnwidth]{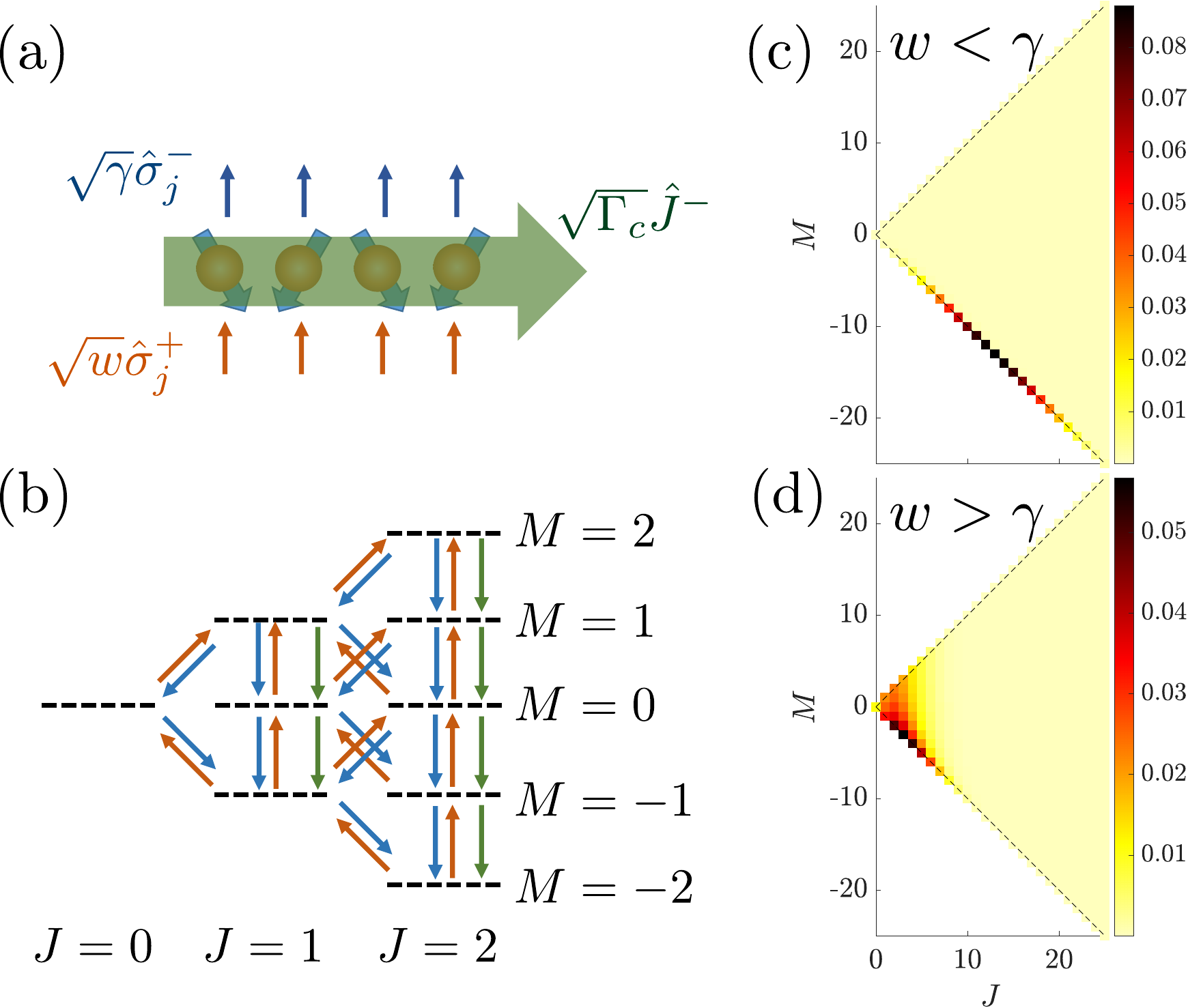}
    \caption{(a-b) Bad cavity laser model illustrated with $N=4$ atoms. The atoms undergo collective decay (green) in the presence of non-collective  pumping (red) and additional non-collective decay (blue). When $w<\gamma+\Gamma_c$, the  phases of the spins are anti-correlated, leading to steady-state subradiant emission. The steady-state density matrix lives in a triangular state space characterized by quantum numbers $J,M$. Collective decay only leads to transitions in the same $J$ manifold, whereas non-collective pumping and decay cause jumps to states in the same as well as adjacent $J$ manifolds. (c-d) Population distribution on the Dicke ladder for $N=100$ atoms, for two states that are approximately equally subradiant (c.f. Eq.~(\ref{eqn:sf})) but on either side of the phase transition.}
    \label{fig:fig1}
\end{figure}

In this Letter, we show that the subradiant regime of a bad cavity laser is in itself a playground for a rich variety of physical phenomena. In particular, we show that the onset of superradiance is preceded by a dissipative phase transition between two distinct types of subradiance. The transition is shown to arise as a consequence of the bounded state space of the collective atomic system. The two subradiant steady states correspond to the population of different regions of this state space (see Fig.~\ref{fig:fig1}). The phase transition is heralded by a non-analytic change in the cavity output power and a discontinuous change in a squeezing parameter. An experimentally attractive feature is the scaling of the output power, which is strongly suppressed and does not increase with atom number $N$ below the critical point but instead scales linearly with $N$ above this point. Near the critical point, we find that the atoms are in a macroscopically entangled state and that the fraction of unentangled atoms is vanishingly small as the number of atoms increases. From the viewpoint of dissipative spin models, this phase transition and the accompanying entanglement is striking because they arise in a model whose governing master equation contains no Hamiltonian terms but only Lindblad dissipators.    

\emph{Model.---} Our system consists of $N$ atoms each with upper and lower levels $\ket{\uparrow}$ and $\ket{\downarrow}$ respectively and a single lossy cavity mode as shown in Fig.~\ref{fig:fig1}(a). The atoms can be modeled using the language of Pauli matrices where $\hSig_j^{-} =  \ket{\downarrow}_j \bra{\uparrow}_j$ ($\hSig_j^{+}=\ket{\uparrow}_j \bra{\downarrow}_j$) is the lowering (raising) operator for atom $j$ and $\hSig_j^{z}=\ket{\uparrow}_j\bra{\uparrow}_j-\ket{\downarrow}_j\bra{\downarrow}_j$ is the population difference between the spin states. The finite lifetime of $\ket{\uparrow}$ causes atoms to emit photons both into free space modes and the cavity mode as they decay to $\ket{\downarrow}$. Emission into free space is characterized by a jump operator $\sqrt{\gamma}\hSig_j^-$ for each atom. Assuming that the atoms are identically coupled to the cavity mode, the emission of a cavity photon is characterized by the jump operator $\sqrt{\Gamma_c}\hat{J}^-$ where $\hat{J}^-=\sum_{j=1}^N \hSig_j^-$ is the collective angular momentum lowering operator. Here, $\Gamma_c=C\gamma$ is the single atom emission rate into the cavity, which is modified by the dimensionless cooperativity parameter $C$. The decay channels are balanced by an effective incoherent pumping of the individual atoms from $\ket{\downarrow}\rightarrow\ket{\uparrow}$ which is represented by a jump operator  
$\sqrt{w}\hSig_j^+$ for each atom. The master equation governing the spin dynamics is therefore given by 
\begin{equation} \label{eqn:MasterEq}
    \rpd \hat{\rho} = \sum_{j=1}^N \hat{\mathcal{D}}\left[\sqrt{w} \hSig_j^{+}\right] \hat{\rho} + \sum_{j=1}^N \hat{\mathcal{D}}\left[\sqrt{\gamma} \hSig_j^{-}\right] \hat{\rho} + \hat{\mathcal{D}}\left[\sqrt{\Gamma_c} \hat{J}^-\right] \hat{\rho},
\end{equation}
where $\hat{\mathcal{D}}[\hat{O}] \hat{\rho} = \hat{O} \hat{\rho} \hat{O}^{\dagger} -  \hat{O}^{\dagger} \hat{O}\hat{\rho}/2 - \hat{\rho} \hat{O}^{\dagger} \hat{O}/2$ is the Lindblad dissipator associated with a jump operator $\hat{O}$. 

\begin{figure*}[!tb]
\centering
    \includegraphics[width=\textwidth]{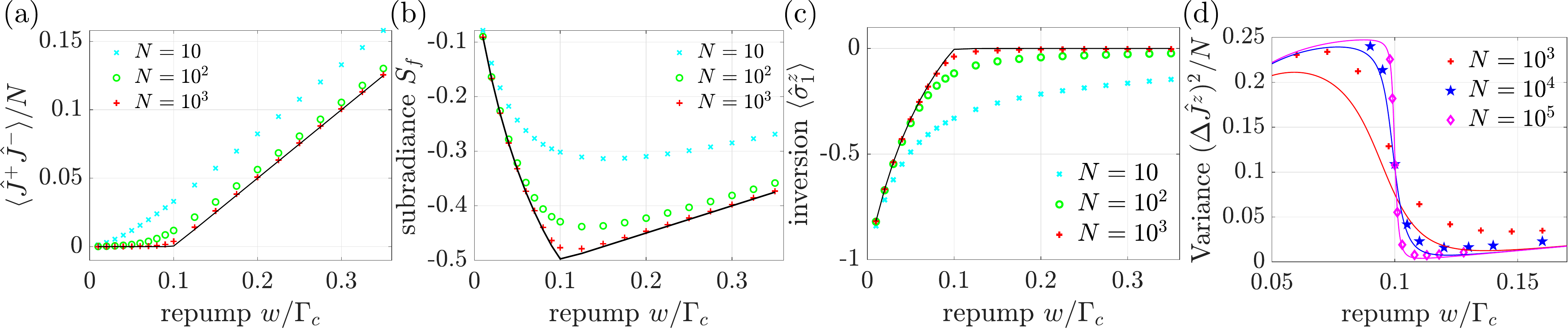}
    \caption{Signatures of the phase transition. (a) Cavity power output per atom, characterized by $\ev{\hat{J}^+\hat{J}^-}/N$, (b) Subradiance factor $S_f$ and (c) Mean atomic inversion, as $w$ is scanned across $\gamma\;(=0.1\Gamma_c)$. Markers show the results from exact diagonalization for various atom numbers. The solid line is the result for $N=10^5$ atoms obtained using second-order cumulant theory. (d) Normalized variance of the total inversion. The solid lines in (d) are computed using third-order cumulant theory for the displayed atom numbers.}
    \label{fig:fig2}
\end{figure*}

This master equation is invariant under permutations of the atomic indices and this symmetry results in a drastic reduction of the Liouville space for the steady-state solution from $4^N$ to $\mathcal{O}(N^3)$ basis states \cite{xu2013PRA,shammahPRA2018}. Furthermore, the master equation also possesses a $U(1)$ symmetry which can be seen by making the transformation $\hSig_j^\pm \rightarrow e^{\pm i\phi}\hSig_j^\pm$ in Eq.~(\ref{eqn:MasterEq}). This additional symmetry reduces the required basis states to $\mathcal{O}(N^2)$. 

A convenient representation of these basis states uses the permutation invariant eigenstates of the $\hat{\mathbf{J}}^2$ and $\hat{J}^z$ operators with respective quantum numbers $J,M$ \cite{suppMat}. Here, we have introduced the collective angular momentum components  $\hat{J}^i=\sum_{j=1}^N\hSig_j^i/2$, $i=x,y,z$, wherein $\hSig_j^x=\hSig_j^++\hSig_j^-$ and $\hSig_j^y=-i(\hSig_j^+-\hSig_j^-)$. The two quantum numbers $J=0,1,2,\ldots,N/2$ (for an even $N$ \footnote{This assumption is purely for computational convenience and for simplifying the presentation. Our results are valid for any $N$ in the $N\gg 1$ regime which is considered here.}) and $M=-J,\ldots,J$ form a discrete, triangular state space for the collective atomic state in Liouville space as shown in Fig.~\ref{fig:fig1}(b). While the two vertices at $J=N/2,M=\pm N/2$ correspond to trivial product states with all spins in $\ket{\uparrow}$ or $\ket{\downarrow}$, the third vertex at $J=0,M=0$ is a highly entangled, subradiant state wherein the atoms are grouped into $N/2$ singlet pairs~\cite{urizar2013PRA}.

In this state space, collective emission leads to a transition with $\Delta M=-1$ within a ladder of constant $J$. While the free space emission and repump of any single atom breaks permutation invariance, the cumulative effect of either of these processes occurring for all atoms preserves this symmetry. Hence, they can be viewed as transitions between different states in this state space with $\Delta M = -1,+1$ respectively. Crucially, these processes couple adjacent $J$ ladders and take the system away from $J=N/2$ which is the initial value when the atomic pseudospins are initialized in a coherent spin state. Closed form expressions for the transition probabilities \cite{zhang2018NJP} enable us to numerically determine the steady-state by exact diagonalization (ED) of a rate matrix \cite{suppMat}. 

\emph{Signatures of the phase transition.---} For repump rates such that $\gamma+\Gamma_c<w<N\Gamma_c$, the system is in the superradiant regime that is characterized by positive inversion and spin-spin correlations $\ev{\hSig_1^z},\ev{\hSig_1^+\hSig_2^-}>0$ \cite{meiser2009PRL}. We now vary $w$ in the weak repump regime $0<w<\gamma+\Gamma_c$ while keeping the values of $\gamma,\Gamma_c$ fixed. We choose $\gamma/\Gamma_c=0.1$, corresponding to $C=10$. We first consider the cavity output power per atom, which is proportional to $\ev{\hat{J}^+\hat{J}^-}/N$, where $\hat{J}^+=(\hat{J}^-)^\dagger$. Figure~\ref{fig:fig2}(a) plots this quantity for different atom numbers as $w$ is scanned across $\gamma$. With increasing system size, we observe signatures of a non-analytic change at $w=\gamma$ that indicates a phase transition. We use second-order cumulant theory to obtain analytical insight into this behavior. Using an expansion in the small parameter $1/N$, we find that the $\mathcal{O}(N^0)$ behavior of $\ev{\hat{J}^+\hat{J}^-}$ is given by \cite{suppMat}
\begin{eqnarray}
\ev{\hat{J}^+\hat{J}^-} = 
\begin{cases}
0 & 0<w<\gamma \\
N\dfrac{w-\gamma}{2\Gamma_c} & \gamma<w<\gamma+\Gamma_c.
\end{cases}
\label{eqn:jpjm}
\end{eqnarray}
For $w<\gamma$, a zero solution at leading order reveals the strong suppression of the cavity output power, which does not grow with $N$ in this regime. On the other hand, the output power grows linearly with $N$ for $w>\gamma$. 

Importantly, this critical point is  distinct from and precedes the onset of superradiance at $w=\gamma+\Gamma_c$. As a result, the collective atomic state is subradiant (with respect to emission into the cavity) in both the phases demarcated by this point. A quantitative measure of the degree of subradiance is the per-atom reduction in the collective emission rate in units of $\Gamma_c$. This subradiance factor $S_f$ is given by  
\begin{equation}
    S_f = \frac{1}{N}\left[\ev{\hat{J}^+\hat{J}^-} - \left(\frac{N}{2} + \ev{\hat{J}^z} \right)\right] = (N-1)\ev{\hSig_1^+\hSig_2^-},
    \label{eqn:sf}
\end{equation}
where $\ev{\hat{J}^+\hat{J}^-}$ describes collective emission and includes the effects of atom-atom correlations, while the second term describes the emission from $N$ uncorrelated atoms. The $J=0,M=0$ singlet state gives the minimum possible value of $S_f=-0.5$ and hence it can be considered the most subradiant state. Remarkably, as shown in Fig.~\ref{fig:fig2}(b), we find that near the critical point $S_f\rightarrow-0.5$ with increasing system size, indicating that the system is highly subradiant on either side of this point and occupies states with $J$ close to zero. 

To understand how these two subradiant phases differ, we plot the population in the $J,M$ states for $N=100$ atoms at two points with similar values of $S_f (\approx -0.37)$ on either side of the critical point (Fig.~\ref{fig:fig1}(c-d)). For $w<\gamma$, the system predominantly occupies the lowest states of each $J$-ladder, i.e., $M=-J$, whereas the value of $J/N \sim \mathcal{O}(1)$ \footnote{A similar population distribution has been previously reported in an atom-cavity system with coherent atomic driving \cite{gegg2018NJP}.}. In contrast, for $w>\gamma$, the system occupies states with vanishing values of $J/N$ whereas all allowed $M$ values are significantly populated. In other words, as $w$ increases, the subradiant system `walks' up the lower boundary of the triangular state space, encounters the vertex at $J=0$ and undergoes a phase transition into a qualitatively different family of  subradiant states. Therefore, the phase transition arises as a result of the closed bottleneck at $J=0$ that reflects the incoming population back into the $J\geq0$ space (see animation \cite{anim}). 

A non-analytic change is also observed in the mean atomic inversion $\ev{\hSig_1^z}=2\ev{\hat{J}^z}/N$, plotted in Fig.~\ref{fig:fig2}(c). We find that $\ev{\hSig_1^z}$ monotonically increases with $w$ for $w<\gamma$ while it is essentially zero (at leading order) for $w>\gamma$ \cite{suppMat}. A further, dramatic evidence for the phase transition is observed in the normalized variance of the collective inversion, given by $(\Delta \hat{J}^z)^2/N$. Figure~\ref{fig:fig2}(d) plots this quantity for $N=10^3,10^4,10^5$ spins. Since $J\ll N/2$ in the critical region, we are able to extend the exact diagonalization (ED) computation to $N \sim 10^5$ by working in a truncated state space with $J_\text{max}\leq 1250$. With increasing atom number, we find strong evidence for a discontinuous jump in this quantity at the critical point. In cumulant theory, we find that this jump in the variance is only reproduced by accounting for third-order cumulants \cite{suppMat}. In particular, we cannot factorize three-atom correlations as  $\ev{\hSig_1^+\hSig_2^-\hSig_3^z}\approx \ev{\hSig_1^+\hSig_2^-}\ev{\hSig_1^z}$. The non-analytic behavior of the inversion and the discontinuity in the variance at the critical point are reminiscent of the behavior of order parameters and susceptibilities in equilibrium phase transitions, but in this system these features manifest in a strongly out-of-equilibrium setting.

\emph{Entanglement.---} The failure of simple mean-field theory to reveal subradiance motivates us to investigate the entanglement properties of the steady state in this regime and in particular near the critical point $w=\gamma$. Since the system occupies states with $J\ll N/2$ near this point, an appropriate entanglement witness is the generalized spin squeezing parameter \cite{toth2007PRL,toth2009PRA} given by 
\begin{eqnarray} \label{SqueezingEq}
\xi^2 = \frac{(\Delta \hat{J}^x)^2+(\Delta \hat{J}^y)^2+(\Delta \hat{J}^z)^2}{N/2},
\end{eqnarray}
where $(\Delta \hat{J}^i)^2=\ev{(\hat{J}^i)^2}-\ev{\hat{J}^i}^2$ is the variance in the spin component $i=x,y,z$. A value of $\xi^2<1$ is sufficient to establish entanglement. Physically, this parameter captures the simultaneous compression of uncertainties in the three angular momentum components and takes the minimum value of $\xi^2=0$ for the macroscopic singlet state with $J=0,M=0$. Furthermore, $\xi^2$ also serves as an upper bound for the fraction of unentangled spins in the system \cite{toth2010NJP}. 

\begin{figure}[!tb]
    \centering
    \includegraphics[width=0.55\columnwidth]{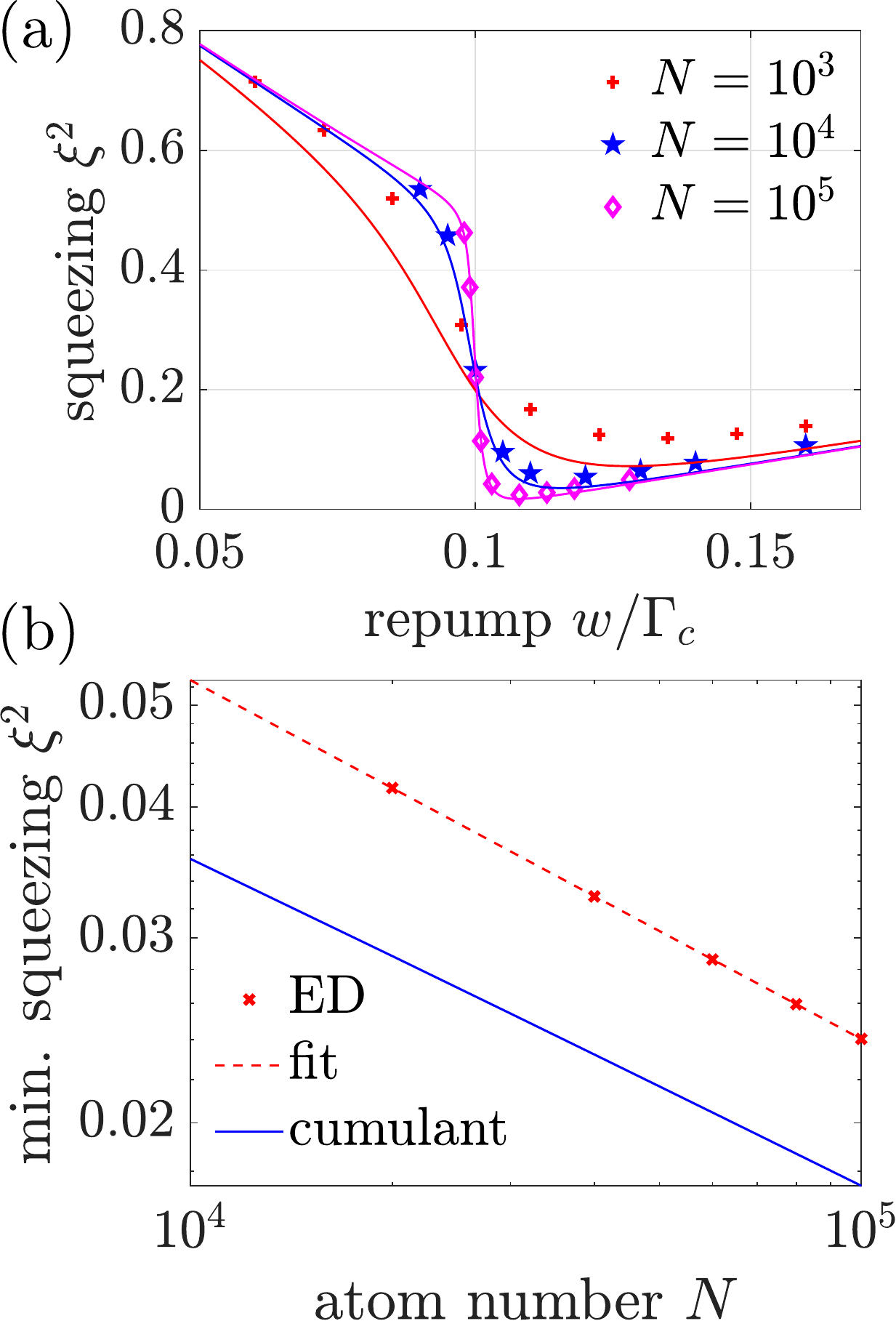}
    \caption{(a) Steady-state squeezing parameter $\xi^2$ as $w$ is scanned across $\gamma\;(=0.1\Gamma_c)$. Markers depict ED results while the lines are obtained from third-order cumulant theory. (b) Minimum $\xi^2$ value extracted from ED and third-order cumulant theory. A fit to the ED data reveals a scaling of $N^{-0.34}$.}
    \label{fig:fig3}
\end{figure}

Figure~\ref{fig:fig3}(a) plots $\xi^2$ as $w$ is varied across $\gamma$. The discontinuity in $(\Delta\hat{J}^z)^2$ also manifests here as a sudden drop in $\xi^2$ near the critical point  that becomes more pronounced with increasing system size. For a finite $N$, the minimum attainable $\xi^2$ decreases with $N$. As shown in Fig.~\ref{fig:fig3}(b), we find a power law scaling $\xi^2\propto N^{-0.34}$ for the minimum value obtained using ED, which is approximately reproduced by the numerical solution of third-order cumulant theory where $\xi^2\propto N^{-0.31}$. This scaling indicates that the fraction of unentangled spins, for which $\xi^2$ is an upper bound, vanishes as $N\rightarrow\infty$. Indeed, in the large $N$ limit, we analytically find that $\xi^2\rightarrow 0$ ($\xi^2\rightarrow 1/2$) as $w\rightarrow\gamma^+$ ($w\rightarrow\gamma^-$) \cite{suppMat}. The subradiant-to-subradiant phase transition is thus characterized by macroscopic entanglement in the atomic ensemble where $O(N)$ atoms are entangled with other atoms.

\emph{Practical considerations.---} Bad cavity lasers based on Raman transitions \cite{bohnet2012Nature} as well as narrow-line optical transitions \cite{norcia2016PRX} can be potentially adapted to observe this transition. Experiments could also be based on cooperative emission from artificial atoms such as NV centers or quantum dots \cite{Angerer:2018,Scheibner:2007}.  Whereas for steady-state superradiance the bad cavity requirement is $\kappa\gg N\Gamma_c$, with $\kappa$  the cavity linewidth and $N\Gamma_c$ the order of the collectively enhanced single-atom emission rate, future studies can explore if this requirement can be relaxed in the subradiant regime where there is no such enhancement. However, similar to the superradiant regime, steady-state subradiance requires the atom-cavity system to satisfy $NC\gg 1$ but operate in the less explored weak pumping limit given by $w\sim\gamma\ll N\Gamma_c$.
Although we have considered the stricter (but achievable \cite{kawasaki2019PRA}) condition $C>1$ in this work, the non-analytic behavior of the inversion and output power is independent of $C$, and the critical scaling of the minimum squeezing with $N$ will also be observable for $C \lesssim 1$, albeit with an exponent of smaller magnitude \cite{suppMat}. However, for $C\ll 1$, the interval $\gamma<w<\gamma+\Gamma_c$ is very small and hence the subradiant-to-subradiant transition is immediately succeeded by the onset of superradiance. We have verified that the mean inversion, output power and the minimum squeezing are robust to $T_2$ dephasing even when $1/T_2 \gtrsim \gamma$ \cite{suppMat}. While $\ev{\hat{J}^+\hat{J}^-}$ can be inferred from the cavity output power, the mean inversion and the variance $(\Delta J^z)^2$ could be measured, for instance, by preparing the steady state and subsequently measuring the population statistics in one of the pseudospin states by detecting the fluorescence from a cycling transition. Alternatively, quantum non-demolition schemes could also be used to measure the latter two observables \cite{hosten2016Nat,cox2016PRL} . The quantities $S_f$ and $\xi^2$ can be estimated by combining these three quantities.  The cavity output can also be used to measure photon bunching via the second-order correlation function $g^{(2)}(0)$, which we find exhibits an abrupt spike at the critical point \cite{suppMat}. 

Identical coupling of the atoms to the cavity mode can be achieved by trapping the atoms at alternate antinodes \cite{hosten2016Nat}. Remarkably, we find that the behavior of the cavity output power and the mean  inversion in the subradiant regime remain unchanged even when the atoms are assumed to be  arbitrarily distributed over a mode wavelength \cite{suppMat}. However, the magnitude of the minimum $S_f$ is reduced because of the modulation by the mode function. This modulation also makes it difficult to infer $S_f$ and $\xi^2$ from measurements of the cavity output and the fluorescence. Importantly, since $\xi^2$ as defined in Eq.~(\ref{SqueezingEq}) does not account for the cavity mode function, it is no longer a suitable entanglement witness since the state can be highly entangled even when $\xi^2>1$. Future work will explore the possibility to construct an entanglement witness that accounts for the cavity mode function.
 
\emph{Conclusion and outlook.---} We have demonstrated that a bad cavity laser undergoes a dissipative phase transition from one subradiant phase to another before the onset of superradiance. Rather than destroying atomic correlations, single atom pumping and decay instead play a central role in generating and maintaining the entangled subradiant states we observe, which, in addition, are also robust to $T_2$ dephasing. Buoyed by recent experiments \cite{Guerin:2016}, subradiance is an exciting frontier with a variety of proposed applications such as ultrafast readouts~\cite{scully2015PRL}, engineering of optical metamaterials~\cite{rui2020Nature}, photon storage~\cite{facchinetti2016PRL,asenjo2017PRX}, quantum state transfer~\cite{guimond2019PRL} and improved quantum metrology~\cite{ostermann2013PRL}, to name but a few. In light of its robust nature, it will be interesting to explore potential applications of steady-state subradiance in quantum information processing, especially considering the features near the critical point such as a vanishing fraction of unentangled spins and an extreme sensitivity of observables to system parameters. From a fundamental perspective, it will be interesting to explore higher-spin models, since the high-dimensional bounded state space presents a greater number of vertices and edges where we may discover dissipative phase transitions that are similar in spirit to the one we have reported here.

\begin{acknowledgments}
We thank Peter Zoller, Walter Hahn, John Cooper, James Thompson and Ana Maria Rey for helpful discussions. A. S. acknowledges support from the European Union$^\prime$s Horizon 2020 research and innovation programme under Grant Agreement No. 731473 (FWF QuantERA via QTFLAG I03769). We also acknowledge support from the NSF AMO Grant No. 1806827; NSF PFC Grant No. 1734006; and the DARPA and ARO Grant No. W911NF-16-1-0576.
\end{acknowledgments}



\begin{thebibliography}{54}
\bibitem [{Note2({\natexlab{b}})}]{anim}%
  \BibitemOpen
  \bibinfo {note} {Media file attached, that shows an animation of the changing
  population distribution in the $J,M$ states as $w$ is scanned across
  $\gamma$.}\BibitemShut {Stop}%
\end{thebibliography}

%

\end{document}


\preprint{APS/123-QED}

\title{Supplemental Material: Subradiant-to-Subradiant Phase Transition in the Bad Cavity Laser}

\author{Athreya Shankar}
\thanks{These authors contributed equally to this work. Corresponding author: A.S.; athreya.shankar@uibk.ac.at}
\affiliation{Center for Quantum Physics, Faculty of Mathematics, Computer Science and Physics, University of Innsbruck, Innsbruck A-6020, Austria}
\affiliation{Institute for Quantum Optics and Quantum Information, Austrian Academy of Sciences, Innsbruck A-6020, Austria}
\author{Jarrod T. Reilly}
\thanks{These authors contributed equally to this work. Corresponding author: A.S.; athreya.shankar@uibk.ac.at}
\affiliation{JILA, NIST, and Department of Physics, University of Colorado Boulder, Boulder, Colorado 80309, USA}
\author{Simon B. J\"ager}
\affiliation{JILA, NIST, and Department of Physics, University of Colorado Boulder, Boulder, Colorado 80309, USA}
\author{Murray J. Holland}
\affiliation{JILA, NIST, and Department of Physics, University of Colorado Boulder, Boulder, Colorado 80309, USA}

\date{\today}

\maketitle

\tableofcontents

\allowdisplaybreaks

\section{Permutation invariant basis states}
A convenient basis to represent pure states of $N$ spin-$1/2$ particles consists of the joint eigenstates of the $\hat{\mathbf{J}}^2$ and $\hat{J}^z$ operators with respective quantum numbers $J$ and $M$ such that:
\begin{equation}
    \begin{aligned}
\hat{\mathbf{J}}^2 \ket{J,M} &= J(J+1) \ket{J,M},\\
\hat{J}^z \ket{J,M} &= M \ket{J,M}.
    \end{aligned}
\end{equation}
Here, the allowed values are $J=0,1,\ldots,N/2$ and $M=-J,-J+1,\ldots,J$ for each value of $J$. For $J<N/2$, the angular momenta of $N$ spin-1/2 particles can in general be combined in multiple ways to yield mutually orthogonal states with the same values of $J,M$. The corresponding degeneracy depends on $J$ and is given by~\cite{mandel1995Optical}  
\begin{equation} \label{JMDegeneracy}
    d_N^J = \frac{N! (2J + 1)}{\left( \frac{N}{2} + J + 1\right)! \left( \frac{N}{2}-J \right)!}.
\end{equation}
Therefore, an auxiliary quantum number $\chi=1,2,\ldots,d_N^J$ is required to distinguish the different orthogonal basis states with the same $J,M$ values.

The master equation considered in this work is invariant under the permutation of particle indices. Because of this permutation symmetry, the steady-state density matrix exists in a restricted Liouville space that is spanned only by permutationally invariant (PI) basis vectors. Each basis has quantum numbers $J,M,M'$ and is uniquely constructed as a uniform statistical mixture given by
\begin{equation}
	\hat{\rho}_{J,M,M'} = \frac{1}{d_N^J}\sum_{\chi=1}^{d_N^J} \ket{N,J,M,\chi}\bra{N,J,M',\chi}.
\end{equation}
Since there are $N/2 +1$ values of $J$ and each of $M,M'$ can take $2J+1$ values for every $J$, the number of PI bases is $ (1/6)(N+1)(N+2)(N+3)\sim\mathcal{O}(N^3)$~\cite{xu2013PRA}. This corresponds to a drastic reduction in the number of bases vectors in Liouville space, which in general is $4^N$ for a collection of $N$ spin-$1/2$ particles. 

In addition, the master equation is invariant under a $U(1)$ transformation which can be verified by the substitution 
\begin{equation}
    \hSig_j^\pm \rightarrow e^{\pm i\phi}\hSig_j^\pm.
\end{equation}
This $U(1)$ symmetry implies that the steady state solution does not have support on PI bases with $M\neq M'$. This results in an additional reduction in the number of bases to $(N+2)^2/4\sim \mathcal{O}(N^2)$. 

\section{Rate equations for determining steady state populations}

\renewcommand{\arraystretch}{2.5}
\setlength{\tabcolsep}{12pt}

\begin{table*}[!tb]
    \centering
    \begin{tabular}{ccc}
    Process &   Transition  &   Rate \\\hline
    Collective decay     &  $(J,M)\rightarrow(J,M-1)$   & $\Gamma_c(J+M)(J-M+1)$  \\
    Individual repump     &   $(J,M)\rightarrow(J,M+1)$   &   $w\left[\dfrac{(N+2)(J-M)(J+M+1)}{4J(J+1)}\right]$   \\
        &   $(J,M)\rightarrow(J-1,M+1)$   &   $w\left[\dfrac{(N+2J+2)(J-M)(J-M-1)}{4J(2J+1)}\right]$   \\
        &   $(J,M)\rightarrow(J+1,M+1)$   &   $w\left[\dfrac{(N-2J)(J+M+1)(J+M+2)}{4(J+1)(2J+1)}\right]$   \\ 
    Individual decay     &   $(J,M)\rightarrow(J,M-1)$   &   $\gamma\left[\dfrac{(N+2)(J+M)(J-M+1)}{4J(J+1)}\right]$   \\
        &   $(J,M)\rightarrow(J-1,M-1)$   &   $\gamma\left[\dfrac{(N+2J+2)(J+M)(J+M-1)}{4J(2J+1)}\right]$   \\
        &   $(J,M)\rightarrow(J+1,M-1)$   &   $\gamma\left[\dfrac{(N-2J)(J-M+1)(J-M+2)}{4(J+1)(2J+1)}\right]$   \\
    Individual dephasing     &   $(J,M)\rightarrow(J,M)$   &   $\dfrac{1}{T_2}\left[\dfrac{(N+2)M}{4J(J+1)}\right]$   \\
        &   $(J,M)\rightarrow(J-1,M)$   &  $\dfrac{1}{T_2}\left[\dfrac{(N+2J+2)(J-M)(J+M)}{4J(2J+1)}\right]$    \\
        &   $(J,M)\rightarrow(J+1,M)$   &   $\dfrac{1}{T_2}\left[\dfrac{(N-2J)(J-M+1)(J+M+1)}{4(J+1)(2J+1)}\right]$   \\
    \end{tabular}
    \caption{Transition rates for different Lindblad terms entering the master equation for the bad cavity laser~\cite{zhang2018NJP}.}
    \label{tab:rates}
\end{table*}

Closed form expressions for transition probabilities between the PI basis states have been previously derived \cite{zhang2018NJP} for the various Lindblad terms constituting master equation (1) of the Main Text. We reproduce these expressions in the form of transition rates in Table~\ref{tab:rates}. We introduce a population vector $\mbf{P}$, whose dynamical evolution is given by 
\begin{eqnarray}
\frac{d}{dt}P_j = \sum_k R_{jk}P_k,
\end{eqnarray}
where $R_{jk}$ represent the elements of a rate matrix $\mathbb{R}$. The indices $j,k$ each take on $(N+2)^2/4$ values corresponding to the total dimension of the PI $U(1)$ symmetric basis set. An off-diagonal element $R_{jk}$ gives the rate of population flow from $k\rightarrow j$ and can be directly read off from Table~\ref{tab:rates}. The diagonal element gives the total rate of population flow out of state $j$ and is thus given by $-\sum_{k \neq j} R_{kj}$. The steady-state population $\mbf{P}_s$ is given by the right eigenvector of $\mathbb{R}$ with eigenvalue $0$, i.e. 
\begin{eqnarray}
\mathbb{R}\mbf{P}_s = 0.
\end{eqnarray}
The populations are normalized according to $\sum_j P_{s,j} = 1$. We note that this normalization is different from the convention adopted in Ref.~\cite{zhang2018NJP} where the degeneracy $d_N^J$ is explicitly factorized out of the state populations. 

\section{Cumulant theory in the subradiant regime}

Our starting point is the  master equation of the bad cavity laser given by  
\begin{equation} \label{eqn:MasterEq_supp}
    \rpd \hat{\rho} = \sum_{j=1}^N \hat{\mathcal{D}}\left[\sqrt{w} \hSig_j^{+}\right] \hat{\rho} + \sum_{j=1}^N \hat{\mathcal{D}}\left[\sqrt{\gamma} \hSig_j^{-}\right] \hat{\rho} + \hat{\mathcal{D}}\left[\sqrt{\Gamma_c} \hat{J}^-\right] \hat{\rho}.
\end{equation}

\subsection{Absence of subradiance in mean-field theory}

The mean-field equations of motion for the expectation values $\ev{\hSig_1^+},\ev{\hSig_1^z}$ are 
\begin{eqnarray}
\frac{d}{dt}\ev{\hSig_1^+} &=& -\frac{\Gamma_+}{2}\ev{\hSig_1^+} + \frac{(N-1)\Gamma_c}{2}\ev{\hSig_1^+}\ev{\hSig_1^z} \nonumber\\
\frac{d}{dt}\ev{\hSig_1^z} &=& -\Gamma_+\ev{\hSig_1^z} + \Gamma_- - 2(N-1)\Gamma_c\abs{\ev{\hSig_1^+}}^2,
\end{eqnarray}
where $\Gamma_\pm=w\pm(\gamma+\Gamma_c)$. These equations admit two solutions given by 
\begin{eqnarray}
\ev{\hSig_1^z} = \Gamma_-/\Gamma_+, &&\;\abs{\ev{\hSig_1^+}} = 0,  \nonumber\\
\ev{\hSig_1^z} = \frac{\Gamma_+}{(N-1)\Gamma_c}, &&\;
\abs{\ev{\hSig_1^+}}^2 = \frac{\Gamma_-}{2(N-1)\Gamma_c} 
- \frac{\Gamma_+^2}{2(N-1)^2\Gamma_c^2}
\label{eqn:mf_sol} \nonumber\\
\end{eqnarray}
For the second solution to represent a physical and nonzero polarization, we require $\abs{\ev{\hSig_1^+}}^2> 0$. Assuming $N \gg 1$,  this condition implies that $w_-<w<w_+$, where 
\begin{eqnarray}
w_- \approx \gamma+\Gamma_c, \;w_+ \approx (N-1)\Gamma_c.
\end{eqnarray}
The two roots, $w_-,w_+$ are respectively the lower and upper bounds for superradiant emission. In this regime, the spin polarization has nonzero magnitude, i.e. $\abs{\ev{\hSig_1^+}}>0$, and the permutation invariance implies that all the spins are polarized along the same direction, giving rise to a classically correlated spin state. 

Instead, for $w<w_-$ or $w>w_+$, the physical solution is the first line of Eq.~(\ref{eqn:mf_sol}) where $\abs{\ev{\hSig_1^+}}=0$, i.e. the spins are unpolarized. The steady state is therefore simply a product state of unpolarized spins. Notably, the subradiant nature of the spin state in the regime  $w<\gamma+\Gamma_c$ does not manifest in mean-field theory.   

\subsection{Second-order cumulant theory} \label{supp:second_order}

We now proceed to study the subradiant regime using second-order cumulant theory. More generally, our analysis is valid in the parameter regime where $w,\gamma,\Gamma_c\ll N\Gamma_c$. Here, the dynamics is described by the following set of equations 
\begin{eqnarray} \label{eqn:second_order_dynamics}
\frac{d}{dt}\ev{\hSig_1^z} &=& -\Gamma_+ \ev{\hSig_1^z} + \Gamma_- - 2(N-1)\Gamma_c \ev{\hSig_1^+\hSig_2^-} \nonumber\\
\frac{d}{dt}\ev{\hSig_1^+\hSig_2^-} &=& -\Gamma_+ \ev{\hSig_1^+\hSig_2^-} + \frac{\Gamma_c}{2}\ev{\hSig_1^z} + \frac{\Gamma_c}{2}\ev{\hSig_1^z\hSig_2^z} \nonumber\\
&&+ (N-2)\Gamma_c\ev{\hSig_1^+\hSig_2^-\hSig_3^z} \nonumber\\
\frac{d}{dt}\ev{\hSig_1^z\hSig_2^z} &=& -2\Gamma_+\ev{\hSig_1^z\hSig_2^z} + 2\Gamma_-\ev{\hSig_1^z} + 4\Gamma_c\ev{\hSig_1^+\hSig_2^-} \nonumber\\
&&-4(N-2)\Gamma_c\ev{\hSig_1^+\hSig_2^-\hSig_3^z}.
\label{eqn:sct_eom}
\end{eqnarray}
To obtain a closed set of equations, we factorize the three-atom moment $\ev{\hSig_1^+\hSig_2^-\hSig_3^z} \approx \ev{\hSig_1^+\hSig_2^-}\ev{\hSig_1^z}$. 

In the subradiant regime, the spin-spin correlation $\ev{\hSig_1^+\hSig_2^-}<0$ and furthermore, its magnitude is $\mathcal{O}(1/N)$. We therefore express the steady-state solution as a hierarchy in orders of $1/N$ such that 
\begin{eqnarray}
\ev{\hat{O}} = \ev{\hat{O}}_0 + \ev{\hat{O}}_1 + \ev{\hat{O}}_2 + \ldots, \end{eqnarray}
where the subscript $n$ denotes $\mathcal{O}(1/N^n)$. In this notation, we immediately have $\ev{\hSig_1^+\hSig_2^-}_0=0$. The leading order solution for the remaining observables are 
\begin{eqnarray}
\ev{\hSig_1^z}_0 = \frac{\Gamma_-}{\Gamma_+} - \frac{2N\Gamma_c}{\Gamma_+}\ev{\hSig_1^+\hSig_2^-}_1, \; \ev{\hSig_1^z\hSig_2^z}_0 = \ev{\hSig_1^z}_0^2.\nonumber\\
\end{eqnarray}
Substituting these expressions into the equation for $\ev{\hSig_1^+\hSig_2^-}$ and retaining only leading order terms, we get 
\begin{eqnarray}
\ev{\hSig_1^z}_0\left(N\ev{\hSig_1^+\hSig_2^-}_1 + \frac{1}{2} +\frac{1}{2}\ev{\hSig_1^z}_0 \right) = 0.
\label{eqn:cum2_quad_sz}
\end{eqnarray}
Equating the two factors to zero results in two solutions given by
\begin{eqnarray}
&&\ev{\hSig_1^z}_0=0,\;\ev{\hSig_1^+\hSig_2^-}_1 = \frac{\Gamma_-}{2N\Gamma_c}, \; \text{for}\; w>\gamma, \nonumber\\
&&\ev{\hSig_1^z}_0 = \frac{w-\gamma}{w+\gamma}, \; \ev{\hSig_1^+\hSig_2^-}_1 = -\frac{w}{N(w+\gamma)}, \; \text{for}\; w<\gamma. \nonumber\\
\label{eqn:sct_leading}
\end{eqnarray}

\subsubsection{Calculation of observables}

\emph{Subradiance factor:} $S_f$ is simply given by $N\ev{\hSig_1^+\hSig_2^-}_1$. Clearly, $S_f$ attains its minimum value of $-0.5$ as $w\rightarrow \gamma$ from either side.

\emph{Cavity output power:} We note that the second factor in Eq.~(\ref{eqn:cum2_quad_sz}) is precisely the leading order expression for $\ev{\hat{J}^+\hat{J}^-}/N$, which is proportional to the power output per atom. This term is exactly zero when $w<\gamma$ but is $\mathcal{O}(1)$ when $w>\gamma$, implying that the phase transition is also marked by a macroscopic emission of photons out of the cavity. Explicitly, we have 
\begin{eqnarray}
\ev{\hat{J}^+\hat{J}^-}_0 = 
\begin{cases}
N\dfrac{w-\gamma}{2\Gamma_c}, & w>\gamma, \\
0, & w<\gamma.
\end{cases}
\label{eqn:jpjm_normalized}
\end{eqnarray}

\emph{Squeezing:} In terms of one-atom and two-atom expectation values, the squeezing parameter $\xi^2$ is 
\begin{eqnarray}
\xi^2 = \frac{3}{2} + 2(N-1)\ev{\hSig_1^+\hSig_2^-} + \frac{N-1}{2}\ev{\hSig_1^z\hSig_2^z} - \frac{N}{2}\ev{\hSig_1^z}^2.  \nonumber\\
\end{eqnarray}
At leading order, $\xi_0^2=0$ since $\ev{\hSig_1^z\hSig_2^z}_0=\ev{\hSig_1^z}_0^2$. At the next order, we get 
\begin{eqnarray}
\xi_1^2 = &&\frac{3}{2} + 2N\ev{\hSig_1^+\hSig_2^-}_1 + \frac{N}{2}\left(\ev{\hSig_1^z\hSig_2^z}_1-2\ev{\hSig_1^z}_0\ev{\hSig_1^z}_1 \right)\nonumber\\
&&-\frac{1}{2}\ev{\hSig_1^z}_0^2.
\label{eqn:xi2_1}
\end{eqnarray}
To evaluate this quantity, we require the next-to-leading order corrections $\ev{\hSig_1^z}_1,\ev{\hSig_1^z\hSig_2^z}_1$. In general, we can always express $\ev{\hSig_1^z}_n.\ev{\hSig_1^z\hSig_2^z}_n$ in terms of $\ev{\hSig_1^+\hSig_2^-}_{n+1}$, which can in turn be iteratively evaluated using  $(d/dt)\ev{\hSig_1^+\hSig_2^-}=0$. Computing these quantities within the scope of second-order cumulant theory, we find that $\xi^2 \rightarrow1/(2+C)$ as $w\rightarrow\gamma$ from either side.
However, the plot of $\xi^2$ versus the repump computed using exact diagonalization shows signs of a discontinuity at $w=\gamma$ and also indicates that the minimum value can be well below this limit. The reason for this inaccuracy is the factorization $\ev{\hSig_1^+\hSig_2^-\hSig_3^z}\approx \ev{\hSig_1^+\hSig_2^-}\ev{\hSig_1^z}$, where we neglect a third-order cumulant that is $\mathcal{O}(1/N^2)$ and hence must be accounted for to correctly compute the next-to-leading order terms that enter the squeezing parameter.

\subsection{Third-order cumulant theory}

We now include the following additional equations 
\begin{widetext}
\begin{eqnarray}
\allowdisplaybreaks
\frac{d}{dt}\ev{\hSig_1^+\hSig_2^-\hSig_3^z} &=& -2(\Gamma_++\Gamma_c)\ev{\hSig_1^+\hSig_2^-\hSig_3^z} + (\Gamma_--\Gamma_c)\ev{\hSig_1^+\hSig_2^-} + \frac{\Gamma_c}{2}\left(\ev{\hSig_1^z\hSig_2^z} + \ev{\hSig_1^z\hSig_2^z\hSig_3^z} \right) \nonumber\\
&&+ (N-3)\Gamma_c\ev{\hSig_1^+\hSig_2^-\hSig_3^z\hSig_4^z} - 2(N-3)\Gamma_c\ev{\hSig_1^+\hSig_2^-\hSig_3^+\hSig_4^-}, \nonumber\\
\frac{d}{dt}\ev{\hSig_1^z\hSig_2^z\hSig_3^z} &=& -3\Gamma_+\ev{\hSig_1^z\hSig_2^z\hSig_3^z} + 3\Gamma_-\ev{\hSig_1^z\hSig_2^z} + 12\Gamma_c\ev{\hSig_1^+\hSig_2^-\hSig_3^z} - 6(N-3)\Gamma_c\ev{\hSig_1^+\hSig_2^-\hSig_3^z\hSig_4^z}.
\label{eqn:tct_eom}
\end{eqnarray}
\end{widetext}

The set of five equations, Eq.~(\ref{eqn:sct_eom}) and Eq.~(\ref{eqn:tct_eom}), are now closed by factorizing the four-atom expectation values according to 
\begin{eqnarray}
\ev{\hSig_1^+\hSig_2^-\hSig_3^+\hSig_4^-} &\approx& 2\ev{\hSig_1^+\hSig_2^-}^2, \nonumber\\
\ev{\hSig_1^+\hSig_2^-\hSig_3^z\hSig_4^z} &\approx& \ev{\hSig_1^+\hSig_2^-}\ev{\hSig_1^z\hSig_2^z} + 2\ev{\hSig_1^+\hSig_2^-\hSig_3^z}\ev{\hSig_1^z} \nonumber\\
&&- 2\ev{\hSig_1^+\hSig_2^-}\ev{\hSig_1^z}^2. 
\end{eqnarray}
We compute numerical results from third-order cumulant theory by finding the steady-state solution to the five equations presented in Eq.~(\ref{eqn:sct_eom}) and (\ref{eqn:tct_eom}). 

To make analytical progress, we assume that a three-atom cumulant will only appear at $\mathcal{O}(1/N^2)$. First, this implies that we can factorize 
\begin{eqnarray}
\ev{\hSig_1^z\hSig_2^z\hSig_3^z} &\approx& 3\ev{\hSig_1^z}\ev{\hSig_1^z\hSig_2^z} - 2\ev{\hSig_1^z}^3
\end{eqnarray}
in the first equation of Eq.~(\ref{eqn:tct_eom}) as long as we  only solve this equation at $\mathcal{O}(1)$ and $\mathcal{O}(1/N)$. The second equation becomes redundant under this factorization. Second, we can also write 
\begin{eqnarray}
\ev{\hSig_1^+\hSig_2^-\hSig_3^z}_0 = 0, \; \ev{\hSig_1^+\hSig_2^-\hSig_3^z}_1 \approx \ev{\hSig_1^+\hSig_2^-}_1\ev{\hSig_1^z}_0, \nonumber\\
\end{eqnarray}
since within the assumed cumulant hierarchy, the factorization should only break down in the $\mathcal{O}(1/N^2)$ term $\ev{\hSig_1^+\hSig_2^-\hSig_3^z}_2$. 

To obtain the next-to-leading order corrections $\ev{\hSig_1^z\hSig_2^z}_1$ and $\ev{\hSig_1^z}_1$ appearing in the squeezing parameter, Eq.~(\ref{eqn:xi2_1}), we solve the coupled set of steady-state equations at $\mathcal{O}(1/N)$. The details of the calculation will be presented elsewhere. Here, we focus on the final result. We find that 
\begin{eqnarray}
    \xi_1^2 = 
    \begin{cases}
    \dfrac{3}{2}\left[\dfrac{w-\gamma}{\Gamma_c}\right], & w>\gamma,\\
    \dfrac{3\gamma-w}{2(w+\gamma)} -\dfrac{1}{2}\left[\dfrac{w-\gamma}{w+\gamma} \right]^2, & w<\gamma.
    \end{cases}
    \label{eqn:xi2_analytic}
\end{eqnarray}
In the limit $w\rightarrow\gamma^+$, we find that $\xi_1^2\rightarrow 0$, i.e. the steady-state is characterized by a vanishing fraction of unentangled spins in the large $N$ limit. On the other hand, as $w\rightarrow\gamma^-$, we find that $\xi_1^2\rightarrow1/2$. The left and right limits at $w=\gamma$ are not equal, implying that the squeezing parameter exhibits a discontinuity at the critical point. The source of the discontinuity is a jump in the value of $\ev{\hSig_1^z\hSig_2^z}_1-2\ev{\hSig_1^z}_0\ev{\hSig_1^z}_1$, which also enters the variance $(\Delta \hat{J}^z)^2$. As a result, we find $(\Delta \hat{J}^z)^2\rightarrow N/4$ as $w\rightarrow\gamma^-$ while $(\Delta \hat{J}^z)^2\rightarrow 0$ for $w\rightarrow\gamma^+$. 

\section{Second-order correlation function}
\begin{figure}[!tb]
    \centering
    \includegraphics[width=0.75\columnwidth]{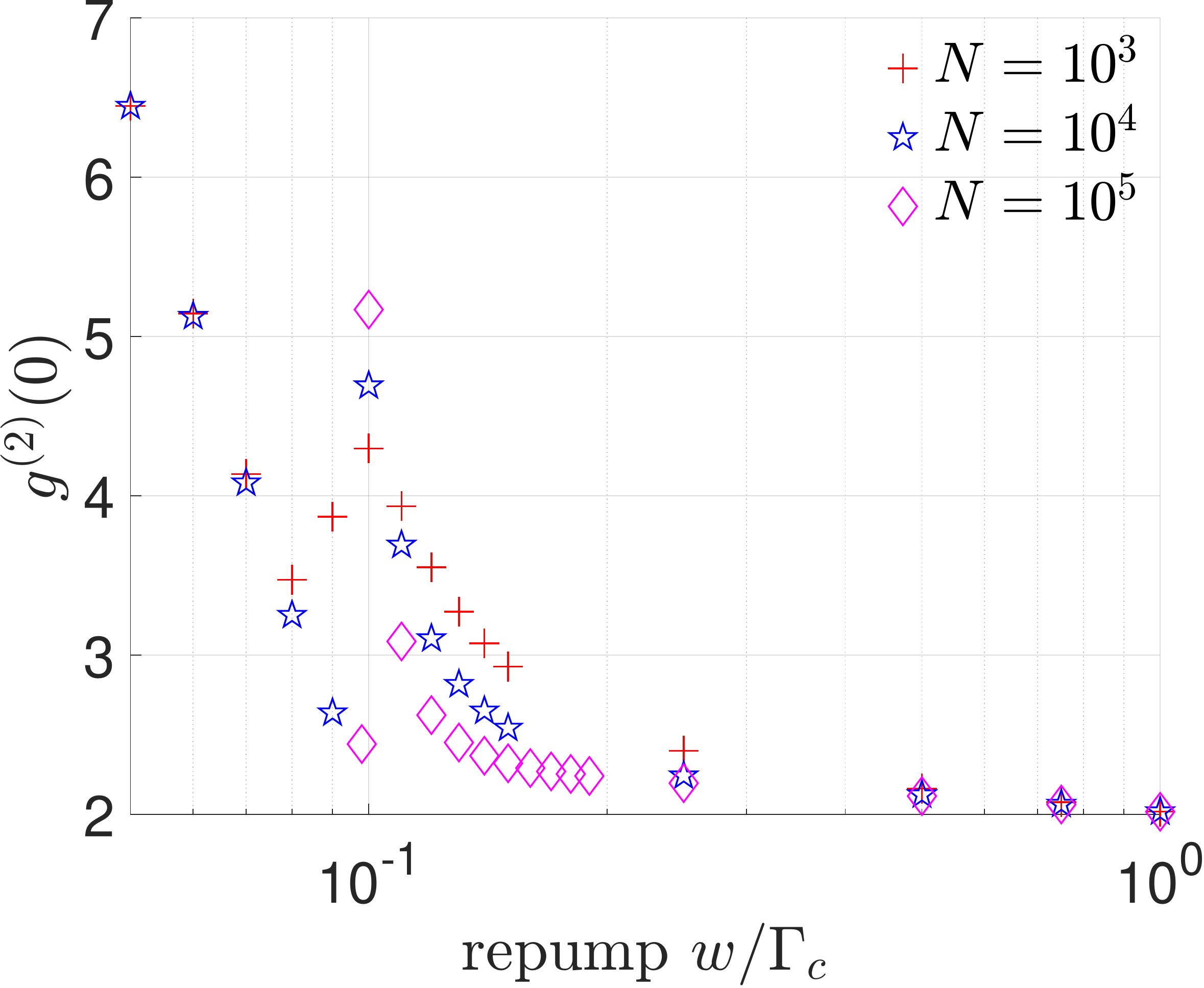}
    \caption{The second-order correlation function at zero time delay $g^{(2)}(0)$, given by Eq.~(\ref{eqn:g2_exp}), as a function of $w$. The markers depict ED results for $N=10^3$, $10^4$, and $10^5$ spins.}
    \label{fig:g2}
\end{figure}

In this section, we study the second-order correlation function at zero time delay
\begin{equation}
    g^{(2)}(0) = \frac{\ev{\hat{a}^{\dagger} \hat{a}^{\dagger} \hat{a} \hat{a}}}{\ev{\hat{a}^{\dagger} \hat{a}}^2},
\end{equation}
which can be expressed in terms of the atomic dipole in the bad-cavity limit as ~\cite{meiser2010PRAIntensity}
\begin{equation}
    g^{(2)}(0) = \frac{\ev{\hat{J}^{+} \hat{J}^{+} \hat{J}^{-} \hat{J}^{-}}}{\ev{\hat{J}^{+} \hat{J}^{-}}^2}.
    \label{eqn:g2_exp}
\end{equation}
Figure~\ref{fig:g2} shows $g^{(2)}(0)$ as $w$ is scanned across $\gamma$ for $N=10^3$ (red, cross), $N=10^4$ (blue, star), and $N=10^5$ (magenta, diamond) atoms. For the cooperativity value $C=10$, the second-order correlation function exhibits an abrupt jump near the critical point $w = \gamma$. For $w>\gamma$, the peak observed near the critical point rapidly decays towards a limiting value of $2$. In the regime $w<\gamma$, we observe that $g^{(2)}(0)$ increases as $w$ decreases. A detailed study of the intensity fluctuations will be the subject of future work. 

\section{Impact of Cooperativity Parameter}

In the Main Text, we choose a large cooperativity of $C = 10$ so that the interval $\gamma < w < \gamma + \Gamma_c$ is discernible from the superradiant regime $w>\gamma+\Gamma_c$. Here, we show that our conclusions are valid even for lower values of $C$.

As Eq.~\eqref{eqn:sct_leading} and Eq.~(\ref{eqn:jpjm_normalized}) demonstrate, the leading order expressions for the inversion and the cavity output power (which is proportional to $\Gamma_c\ev{\hat{J}^+\hat{J}^-}$) are independent of $C$. The invariance of the latter observable is experimentally appealing because the phase transition can be detected by a macroscopic change in the cavity output power irrespective of the value of $C$. From Eq.~(\ref{eqn:xi2_analytic}), the squeezing parameter is independent of $C$ for $w<\gamma$ while being proportional to $1/C$ when $w>\gamma$. Nevertheless, $\xi^2\rightarrow 0$ as $w\rightarrow\gamma^+$ in the large $N$ limit. We therefore conclude that the system in both intervals $w<\gamma$ and $\gamma < w < \gamma + \Gamma_c$ continues to exhibit two distinct forms of subradiant behavior and undergoes a phase transition at $w=\gamma$ for any value of $C$, although the latter regime may be difficult to examine when $C \ll 1$ as the critical point would be almost immediately followed by the onset of superradiance.

\begin{figure}[!tb]
    \centering
    \includegraphics[width=0.75\columnwidth]{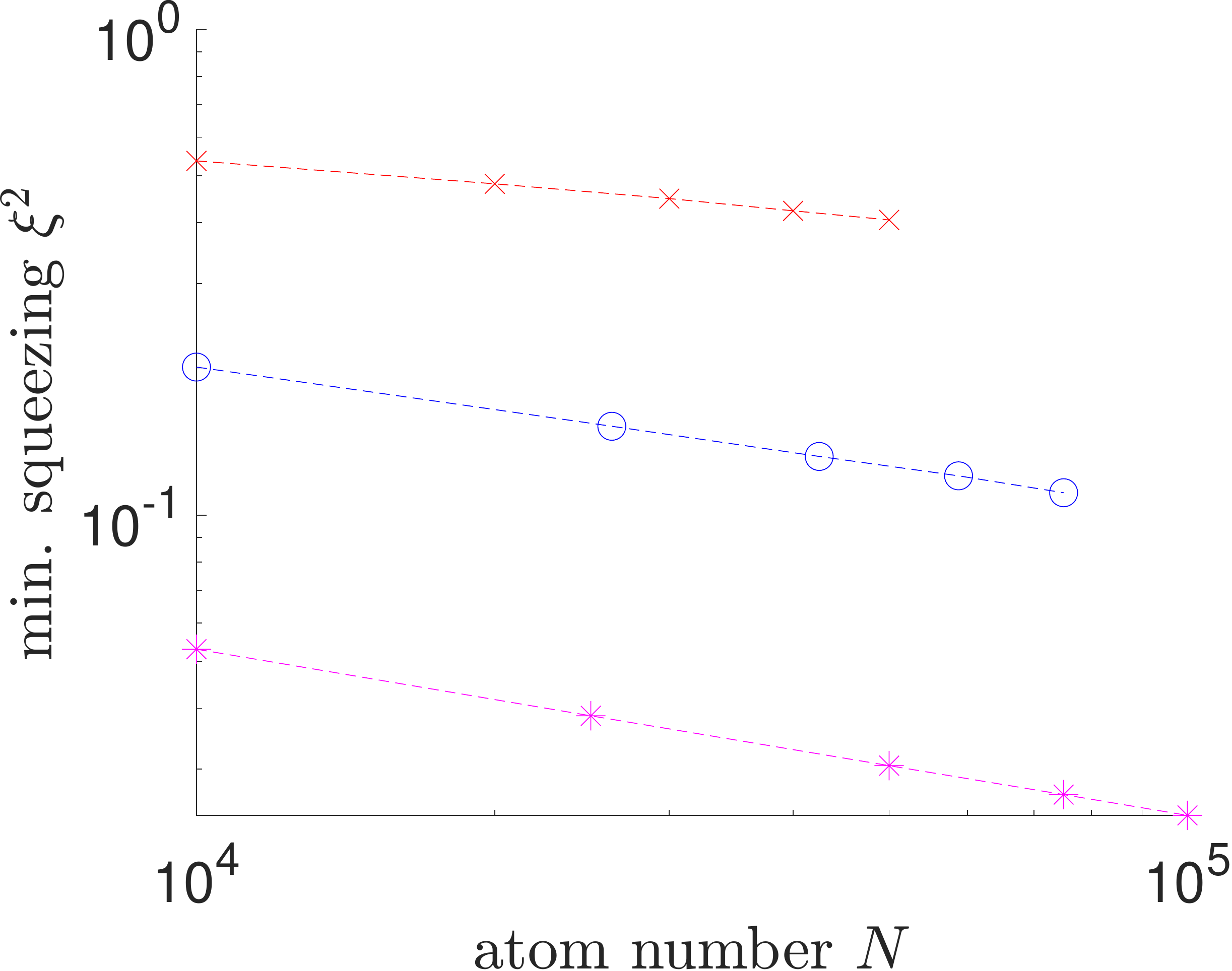}
    \caption{ED results of the minimum squeezing parameter $\xi^2$ as a function of atom number for $C=0.1$ (red, cross), $C=1$ (blue, circle), and $C=10$ (magenta, asterisk). The corresponding curve fits (dashed lines) reveal a scaling of $N^{-0.17}$, $N^{-0.3}$, and $N^{-0.34}$, respectively.}
    \label{fig:impact_C}
\end{figure}

We now demonstrate that a critical scaling of the squeezing parameter can be observed even at smaller $C$ values.
Figure~\ref{fig:impact_C} depicts the minimum value of $\xi^2$ for a varying number of atoms. The data correspond to $C=0.1$ (red, cross), $C=1$ (blue, circle), and $C=10$ (magenta, asterisk). The dashed lines are the corresponding curve fits which reveal that the scaling with $N$ is similar for $C \gtrsim 1$, but is noticeably reduced for smaller values of $C$.

\section{Impact of Dephasing}

\begin{figure}[!tb]
    \centering
    \includegraphics[width=0.35\columnwidth]{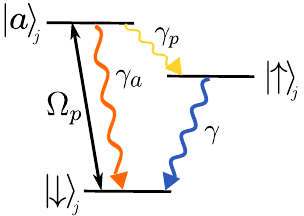}
    \caption{Dephasing as a byproduct of repumping. Atom $j$ is coupled from the ground state $\ket{\downarrow}_j$ to an auxiliary state $\ket{a}_j$ by a laser with effective Rabi frequency $\Omega_p$. The rapid decay $\ket{a}_j\rightarrow\ket{\uparrow}$ leads to an effective repump process from  $\ket{\downarrow}_j\rightarrow\ket{\uparrow}_j$ at rate $w = \Omega_p^2 \gamma_p/[\Gamma(\gamma_p + \gamma_a)]$, where  $\Gamma$ accounts for  $\gamma_p,\gamma_a$ and a possible non-zero linewidth of the coupling laser~\cite{cooper1981PRA}. On the other hand,  the decay $\ket{a}_j\rightarrow\ket{\downarrow}_j$ does not change the pseudospin state but destroys coherence between the spin states. The strength of this dephasing can be characterized by a rate $1/T_2 = \Omega_p^2 \gamma_a/\left[4\Gamma(\gamma_p + \gamma_a)\right]$.}
    \label{fig:pumping}
\end{figure}

In this Section, we probe the robustness of the subradiant-to-subradiant phase transition to individual dephasing of the atomic dipoles. Dephasing can arise as a byproduct of the repump process (see Fig.~\ref{fig:pumping}) or due to ambient effects such as atom-atom collisions and photon scattering from the optical lattice used for trapping.

The master equation including dephasing has the form 
\begin{eqnarray} \label{supp:DephasingMasterEq}
    \rpd \hat{\rho} = &&\sum_{j=1}^N \hat{\mathcal{D}}\left[\sqrt{w} \hSig_j^{+}\right] \hat{\rho} + \sum_{j=1}^N \hat{\mathcal{D}}\left[\sqrt{\gamma} \hSig_j^{-}\right] \hat{\rho} \nonumber\\
    &&+ \hat{\mathcal{D}}\left[\sqrt{\Gamma_c} \hat{J}^-\right] \hat{\rho} + \sum_{j=1}^N \hat{\mathcal{D}}\left[\frac{1}{\sqrt{T_2}} \hSig_j^{z}\right] \hat{\rho},
\end{eqnarray}
where $1/T_2$ is the dephasing rate. 
We can gain insight on the system's steady state by examining the second-order cumulant equations.
The only modification to Eq.~\eqref{eqn:second_order_dynamics} is an increased decoherence rate for the spin-spin correlations:
\begin{eqnarray}
    \frac{d}{dt} \ev{\hSig_1^+ \hSig_2^-} = &&- \left( \Gamma_+ + \frac{4}{T_2} \right) \ev{\hSig_1^+ \hSig_2^-} 
    + \frac{\Gamma_c}{2}\left[ \ev{\hSig_1^z} + \ev{\hSig_1^z \hSig_2^z}\right]
    \nonumber\\
    && + (N-2) \Gamma_c \ev{\hSig_1^+ \hSig_2^-} \ev{\hSig_1^z} .
\end{eqnarray} 
However, for $1/T_2 \ll N\Gamma_c$, this modified term still remains $\mathcal{O}(1/N)$ and hence does not contribute to the leading order expressions for the inversion or spin-spin correlation reported in Eq.~\eqref{eqn:sct_leading}. Therefore, the two subradiant regimes can still be distinguished on the basis of the mean inversion and the cavity output power, which remain unchanged at leading order even in the presence of appreciable levels of dephasing. By similarly inspecting the equations in third-order cumulant theory, we find that the leading order expressions for the squeezing parameter in the two subradiant phases are also unchanged provided $1/T_2 \ll N\Gamma_c$.

\begin{figure}[!tb]
    \centering
    \includegraphics[width=0.75\columnwidth]{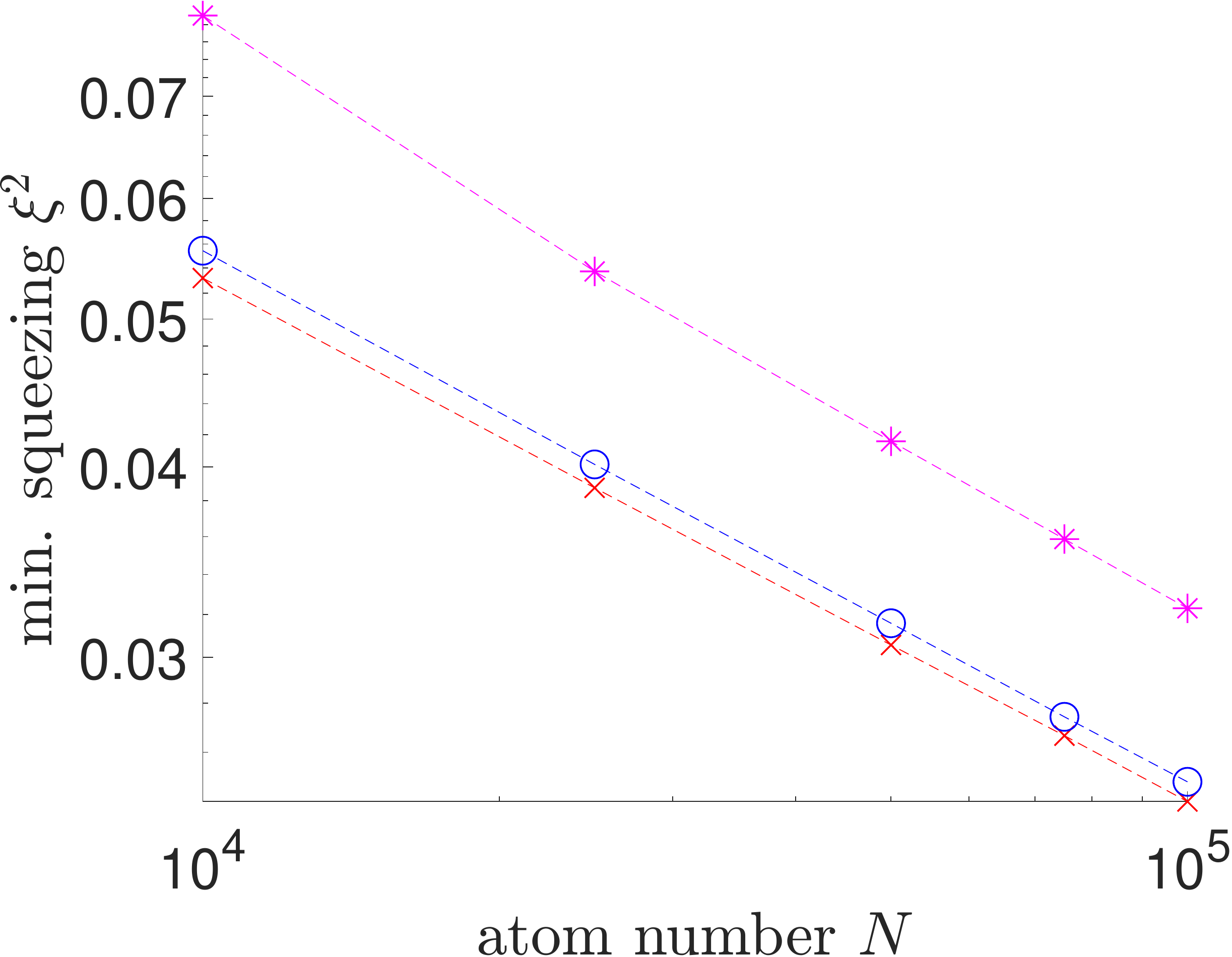}
    \caption{The markers represent the minimum squeezing parameter $\xi^2$ for three values of $\alpha \equiv 1/(wT_2)$: $\alpha = 0.1$, (red, cross), $\alpha=1$ (blue, circle), and $\alpha=10$ (magenta, asterisk).
    The dashed lines depict the corresponding curve fits which demonstrate the robustness of the critical scaling to $T_2$ dephasing.}
    \label{fig:dephasing}
\end{figure}

We now show using ED that the critical scaling of $\xi^2$ is also largely unaffected by dephasing. We model dephasing to arise as a byproduct of the repump process, as depicted in Fig.~\ref{fig:pumping}. In that case,  $1/T_2\propto w$ and we can introduce a dimensionless parameter $\alpha \equiv 1/(w T_2)$ to quantify the strength of dephasing. In Fig.~\ref{fig:dephasing}, we plot the minimum of $\xi^2$ as a function of atom number for $1/T_2$ varying over three orders of magnitude by choosing $\alpha = 0.1$, (red, cross), $\alpha=1$ (blue, circle), and $\alpha=10$ (magenta, asterisk). We make a curve fit for each set of ED results, which is depicted by the dashed lines. We find a similar scaling of the minimum $\xi^2$ for all three values of $\alpha$, demonstrating that the critical scaling is very robust to $T_2$ dephasing. Furthermore, Fig.~\ref{fig:dephasing} demonstrates that the minimum value of $\xi^2$ remains of the same order as we vary the strength of dephasing. The robustness of the observables to dephasing is a major advantage that facilitates experimental observations of the phase transition and may potentially allow for more freedom in the choice of atomic transition.


\section{Inhomogeneous atom-cavity coupling}

\begin{figure}[!tb]
    \centering
    \includegraphics[width=0.6\columnwidth]{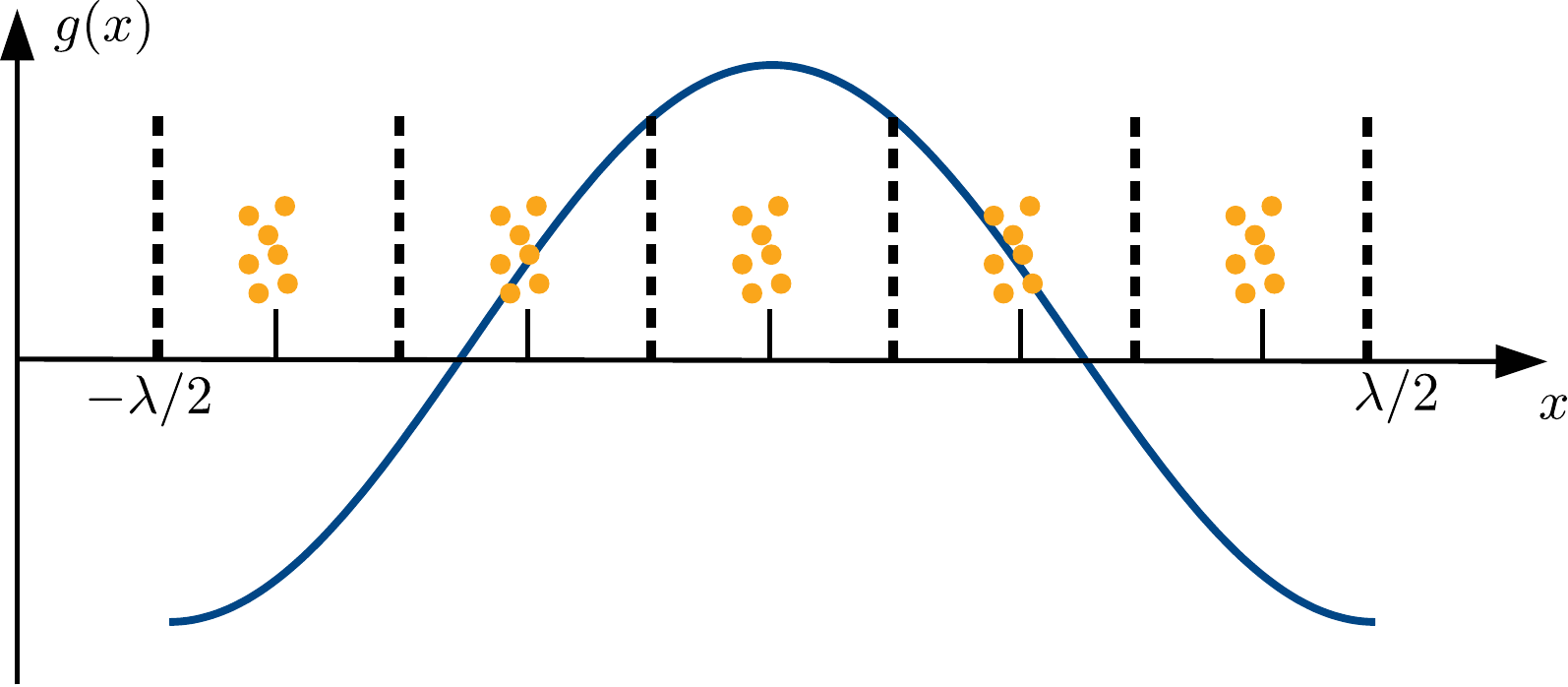}
    \caption{The inhomogeneous coupling of atoms to the cavity mode is modeled by dividing the interval corresponding to one wavelength into $M$ bins and distributing the atoms into these bins according to a given distribution.}
    \label{fig:inhom_bins}
\end{figure}

In this Section, we will investigate the situation where the atoms are not trapped at the antinodes but are spread over the cavity mode function, given by $g(x)=g_0 \cos(2\pi x/\lambda)$, where $\lambda$ is the wavelength. As illustrated in Fig.~\ref{fig:inhom_bins}, we divide the region $-\lambda/2\leq x \leq \lambda/2$ into $M$ bins and assume that the atoms are positioned at the bin centers. We suppose that the number of atoms at $x_m$ is $N_m$, such that $\sum_m N_m = N$. After eliminating the cavity mode, the master equation for the bad cavity laser is now of the form
\begin{eqnarray}
    &&\rpd \hat{\rho} = \sum_{j=1}^N \hat{\mathcal{D}}\left[\sqrt{w} \hSig_j^{+}\right] \hat{\rho} + \sum_{j=1}^N \hat{\mathcal{D}}\left[\sqrt{\gamma} \hSig_j^{-}\right] \hat{\rho} \nonumber\\
    &&+ \sum_{n,m=1}^{M}\sum_{i_n=1}^{N_n}\sum_{j_m=1}^{N_m}\frac{\Gamma_c^{n,m}}{2}\left(2\hSig_{i_n}^-\hat{\rho}\hSig_{j_m}^+ - \hSig_{j_m}^+\hSig_{i_n}^-\hat{\rho} - \hat{\rho} \hSig_{j_m}^+\hSig_{i_n}^- \right). \nonumber\\
    \label{eqn:me_inhom}
\end{eqnarray}
In the last term, the indices $i_n,j_m$ respectively sum over atoms in bins $n,m$. The atom-atom coupling constant now varies depending on the bin index and is given by 
\begin{eqnarray}
    \Gamma_c^{n,m} = \Gamma_c \cos\left(\theta_n \right) \cos\left(\theta_m \right),
\end{eqnarray}
where $\theta_m = 2\pi(m-1/2)/M$ and $\Gamma_c=g_0^2/\kappa$ for a cavity with decay rate $\kappa$. 

The master equation~(\ref{eqn:me_inhom}) is not permutation invariant and hence an exact diagonalization is no longer tractable. Instead, we use second-order cumulant theory to investigate the effect of inhomogeneous atom-cavity coupling. The relevant variables are $\ev{\hSig_{1,k}^z}, \ev{\hSig_{1,k}^z\hSig_{2,q}^z}$ and  $\ev{\hSig_{1,k}^+\hSig_{2,q}^-}_s = \left(\ev{\hSig_{1,k}^+\hSig_{2,q}^-}+ \ev{\hSig_{2,q}^+\hSig_{1,k}^-}\right) /2$, where $k,q=1,\ldots,M$. The equations for the expectation values are given by 
\begin{widetext}
\begin{eqnarray}
    \frac{d}{dt}\ev{\hSig_{1,k}^z} &=& -(w+\gamma+\Gamma_c^{k,k})\ev{\hSig_{1,k}^z} 
    +(w-\gamma-\Gamma_c^{k,k})
    -2\sum_{m}(N_m-\delta_{k,m})\Gamma_c^{k,m}\ev{\hSig_{1,k}^+\hSig_{2,m}^-}_s \nonumber\\
    \frac{d}{dt}\ev{\hSig_{1,k}^+\hSig_{2,q}^-}_s &=& -\left(w+\gamma+\frac{\Gamma_c^{k,k}+\Gamma_c^{q,q}}{2}\right) \ev{\hSig_{1,k}^+\hSig_{2,q}^-}_s
    +\frac{\Gamma_c^{k,q}}{2}\ev{\hSig_{1,k}^z\hSig_{2,q}^z}
    +\frac{\Gamma_c^{k,q}}{4}\left(\ev{\hSig_{1,k}^z}+\ev{\hSig_{2,q}^z}\right) \nonumber\\
    &&+\frac{1}{2}\sum_{m}(N_m-\delta_{k,m}-\delta_{q,m})\left(\Gamma_c^{k,m}\ev{\hSig_{1,m}^+\hSig_{2,q}^-}_s\ev{\hSig_{1,k}^z}
    +\Gamma_c^{m,q}\ev{\hSig_{1,k}^+\hSig_{2,m}^-}_s\ev{\hSig_{1,q}^z}\right) \nonumber\\
    \frac{d}{dt}\ev{\hSig_{1,k}^z\hSig_{2,q}^z} &=& -2\left(w+\gamma+\frac{\Gamma_c^{k,k}+\Gamma_c^{q,q}}{2}\right) \ev{\hSig_{1,k}^z\hSig_{2,q}^z}
    + (w-\gamma-\Gamma_c^{k,k})\ev{\hSig_{2,q}^z} + (w-\gamma-\Gamma_c^{q,q})\ev{\hSig_{1,k}^z} \nonumber\\
    &&-2\sum_{m}(N_m-\delta_{k,m}-\delta_{m,q})\left(\Gamma_c^{k,m}\ev{\hSig_{1,k}^+\hSig_{2,m}^-}_s\ev{\hSig_{1,q}^z} 
    + \Gamma_c^{m,q}\ev{\hSig_{1,m}^+\hSig_{2,q}^-}_s\ev{\hSig_{1,k}^z} \right) \nonumber\\
    &&+4\Gamma_c^{k,q}\ev{\hSig_{1,k}^+\hSig_{2,q}^-}_s.
    \label{eqn:inhom_eom}
\end{eqnarray}
\end{widetext}

We consider three observables, namely the total inversion, the subradiance factor $S_f$ and the power output from the cavity which is proportional to $\Gamma_c \mathcal{P}$, where $\mathcal{P}=\ev{\hat{J}^+\hat{J}^-}$ in the case of uniform light-atom coupling. The expressions for these observables are now modified to account for the cavity mode function as
\begin{eqnarray}
    \ev{\hat{J}^z} &=& \sum_{m=1}^M N_m \ev{\hSig_{1,m}^z}, \nonumber\\
   \Gamma_c \mathcal{P} &=& \sum_{m=1}^M \Gamma_c^{m,m}\frac{N_m}{2}(1+\ev{\hSig_{1,m}^z}) \nonumber\\
   &&+ \sum_{m,n=1}^M \Gamma_c^{m,n}(N_m N_n - N_m\delta_{m,n}) \ev{\hSig_{1,m}^+\hSig_{2,n}^-}_s, \nonumber\\
   S_f &=& \frac{1}{N\Gamma_c}\left(\Gamma_c\mathcal{P} - \sum_{m=1}^M \Gamma_c^{m,m}\frac{N_m}{2}(1+\ev{\hSig_{1,m}^z})\right). \nonumber\\
   \label{eqn:inhom_obs}
\end{eqnarray}

We derive analytic expressions from Eq.~(\ref{eqn:inhom_eom}) by using an order-by-order expansion similar to that used in the permutation symmetric case. We assume $\ev{\hSig_{1,k}^z\hSig_{2,q}^z}_0 = \ev{\hSig_{1,k}^z}_0\ev{\hSig_{2,q}^z}_0$ since this relation holds in the permutation symmetric case. We then define 
\begin{eqnarray}
    X_k = \frac{1}{N}\sum_m N_m \cos\theta_m\ev{\hSig_{1,k}^+\hSig_{2,m}^-}_{s,1}, 
\end{eqnarray}
and find two possible solutions given by 
\begin{eqnarray}
    &&\ev{\hSig_{1,k}^z}_0 = 0, \; X_k = \frac{w-\gamma-\Gamma_c\cos^2\theta_k}{2N\Gamma_c\cos\theta_k}, \; \text{for} \; w>\gamma \nonumber\\
    &&\ev{\hSig_{1,k}^z}_0 = \frac{w-\gamma}{w+\gamma}, \; X_k = -\frac{w\cos\theta_k}{N(w+\gamma)}, \; \text{for} \; w<\gamma. \nonumber\\
\end{eqnarray}
From these solutions, it is evident that the mean inversion remains the same even when the atoms are arbitrarily distributed over the cavity mode function. Simple algebra shows that, remarkably, even the cavity output power is independent of the distribution and is given by Eq.~(2) of the Main Text (multiplied by the rate $\Gamma_c$). Therefore, the phase transition can still be observed as a non-analytic behavior in either of these quantities. The subradiance factor $S_f$, however, depends on the atomic distribution and is given by 
\begin{eqnarray}
    S_f = 
    \begin{cases}
    \dfrac{w-\gamma-\Gamma_c\overline{\cos^2\theta}}{2\Gamma_c}, \; w>\gamma, \nonumber\\
    -\dfrac{w\overline{\cos^2\theta}}{w+\gamma}, \; w<\gamma,
    \end{cases}
\end{eqnarray}
where the overbar denotes the average over the spatial distribution of atoms. As a result, the minimum value of $S_f$ is $-1/4$ when the atoms are uniformly distributed over a wavelength whereas it is $-1/2$ when the atoms are trapped at the antinodes. We have verified the validity of these analytic results by comparing with the numerical steady-state solution to Eq.~(\ref{eqn:inhom_eom}). To verify the accuracy of these expressions, we have numerically solved for the steady state of Eq.~(\ref{eqn:inhom_eom}) taking the case of $N=10^4$ atoms uniformly distributed over either $M=25$ or $M=40$ bins that span the cavity wavelength. We find the numerical results to be in excellent agreement with the above expressions. 

As already mentioned in the Main Text, the squeezing parameter $\xi^2$ defined in Eq.~(4) of the main text does not account for the cavity mode function and hence can be significantly larger than $1$ even when the atoms are in a highly entangled state.



%